\documentclass[journal,twoside,web]{./cls/lcss/ieeecolor}

\usepackage{./cls/lcss/lcsys}
\usepackage{cite}
\usepackage{amsmath,amssymb,amsfonts}
\usepackage{algorithmic}
\usepackage{graphicx}
\usepackage{textcomp}
\usepackage{mathtools}

\usepackage[hidelinks]{hyperref} 
\hypersetup{
    pdftitle={ReLU ANN Feedback linearization}
    }
\newcommand{\be}{\begin{equation}}
\newcommand{\ee}{\end{equation}}

\newcommand{\bald}{\begin{aligned}}
\newcommand{\eald}{\end{aligned}}

\newcommand{\bbm}{\begin{bmatrix}}
\newcommand{\ebm}{\end{bmatrix}}

\newcommand{\bu}{\boldsymbol{u}}

\newcommand{\bx}{\boldsymbol{x}}
\newcommand{\bv}{\boldsymbol{v}}

\newcommand{\R}{\mathbb{R}}

\newcommand{\bb}{\boldsymbol{b}}

\newcommand{\bz}{\boldsymbol{z}}
\newcommand{\by}{\boldsymbol{y}}
\newcommand{\bW}{\boldsymbol{W}}

\newcommand{\bep}{\boldsymbol{\epsilon}}

\newcommand{\umax}[1]{{#1}_{max}}

\newtheorem{rem}{Remark}
\newtheorem{prop}{Proposition}

\newcommand{\QEDA}{\hfill\ensuremath{\square}}

\usepackage{tikz}
\usepackage{color}
\usepackage{multirow}
\usepackage{amssymb}
\usepackage{gensymb}
\usepackage{tabularx}
\usepackage{extarrows}
\usetikzlibrary{fadings}
\usetikzlibrary{patterns}
\usetikzlibrary{shadows.blur}
\usetikzlibrary{shapes}

\usepackage{pgfplots}
\pgfplotsset{width=10cm,compat=1.9}

\definecolor{midnightGreen}{RGB}{11,85,99}
\definecolor{celestialBlue}{RGB}{82, 153, 211}

\usepackage{cases} 
\usepackage[ruled,norelsize,linesnumbered]{algorithm2e}
\newcommand{\figlink}{./Tex_code/tikz_code_fig/cdc2024}
\newcommand{\inte}[2]{\mathcal{I}_{#1,#2}}

\begin{document}
\title{On the constrained feedback linearization control based on the MILP representation\\of a ReLU-ANN}
\author{Huu-Thinh Do, Ionela Prodan
\thanks{
Huu-Thinh Do thanks his colleagues Duc-Tri Vo and Julien Soulé from LCIS for the fruitful discussions on ANNs.
The work of Ionela Prodan was supported is funded by La Région Auvergne-Rhône-Alpes, Pack Ambition Recherche 2021 - PlanMAV, RECPLAMALCIR and Ambition Internationale 2023, Horizon-TA C7H-REG24A10. She also benefits from the support of the FMJH Program PGMO and from the support to this program from EDF.
}
\thanks{Huu-Thinh Do and Ionela Prodan are with Univ. Grenoble Alpes, Grenoble INP$^\dagger$, LCIS, F-26000, Valence, France.
(email:  \{huu-thinh.do,ionela.prodan\}@lcis.grenoble-inp.fr).
\newline
$^\dagger$Institute of Engineering and Management Univ. Grenoble Alpes.  }
}

\maketitle
\thispagestyle{empty}

\begin{abstract}
In this work, we explore the efficacy of rectified linear unit artificial neural networks in addressing the intricate challenges of convoluted constraints arising from feedback linearization mapping. Our approach involves a comprehensive procedure, encompassing the approximation of constraints through a regression process. Subsequently, we transform these constraints into an equivalent representation of mixed-integer linear constraints, seamlessly integrating them into other stabilizing control architectures. The advantage resides in the compatibility with the linear control design and the constraint satisfaction in the model predictive control setup, even for forecasted trajectories. Simulations are provided to validate the proposed constraint reformulation.
\end{abstract}


\section{Introduction}
\label{sec:Intro}
\IEEEPARstart{T}{he} feedback linearization (FL) technique, although being investigated for decades {\cite{isidori1985nonlinear}}, still attracts considerable attention in both industry and research community thanks to its connection from nonlinear design to the well-developed control synthesis for linear systems. 
The method's popularity grows even stronger when the dynamic feedback linearizability is connected to the notion of flat systems on a differential geometry foundation \cite{fliess1993differentially}, giving rise to many applications in applied engineering \cite{greeff2020exploiting,vincentECC23,freire2023flatness}. The general idea of FL includes
a variable transformation that reveals the linear structure of the original nonlinear dynamics in new coordinates, reducing the control design to closing the loop for the linearized dynamics. 
However, the advantage does not generally hold when operating constraints are considered (such as input and state constraints). This is because, while the nonlinearity is cancelled in the dynamics, the constraints are rendered nonlinear and state-dependent, which poses challenges in both theoretical analysis and real-time implementation.
\begin{figure}[tb]
    \centering

\resizebox{0.45\textwidth}{!}{\input{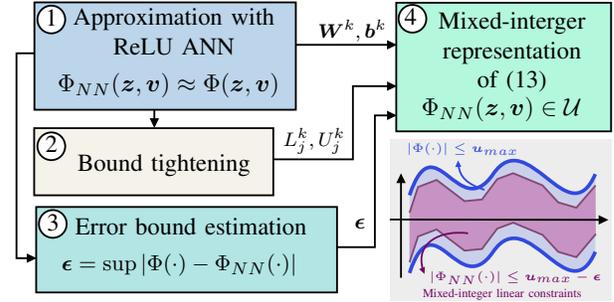}}
    \caption{Constraint characterization: from ReLU-ANN to MILP.}
    \label{fig:MIP_form_relu}
\end{figure}

To address these problems, several studies have been conducted with different approaches. 
On one hand, to provide rigorous stability guarantees by exploiting the constraints' structural characteristics (e.g., shapes, convexity, or explicit bounds), specific investigations have been carried out: for inverted pendulum with nested control \cite{ibanez2008controlling}, linear matrix inequality-based design for wheeled mobile robots \cite{tiriolo2023set}, or polytope-based constraint characterization for quadcopters \cite{nguyen2024notes,thinhECC23,greeff2018flatness}.
Although these methods were proven effective, the problem of generally converting the convoluted constraints into an online, tractable program remains open.

On the other hand, with the ability to handle constraints, Model Predictive Control (MPC) constitutes a promising candidate for navigation in the new coordinates. However, although the prediction model is linear, the constraints enforced are not trivial to solve since, over a prediction horizon, they are defined implicitly by the decision variables (the predicted states and inputs), creating a computationally intensive program online. To relax this setting and achieve real-time operability, approximation-based solutions have been proposed, creating control laws with the complexity of quadratic programs (QP).
The most common technique computes the actual constraint via state feedback and extends it constantly to the whole prediction horizon
\cite{nevistic1994feasible,kandler2012differential}. In a less conservative manner, predictions from the previous time step are employed to approximate the exact actual constraints \cite{kurtz1998feedback}. 
While both of these solutions demonstrate computational efficiency, they pose challenges in terms of stability analysis development \cite{kurtz1998feedback}. This issue persists even under the assumption of online feasibility, as the proof of constraint satisfaction is limited to the applied input rather than the entire predicted state or input trajectory.
Recently, probabilistic approaches have been used to learn the constraints via Gaussian processes \cite{hall2023differentially,greeff2021learning}, leveraging the affine structure in single-input flat systems to formulate a controller as a second-order cone program.

In our pursuit of constructing a robust approximation of the constraints and rendering the online routine tractable, this study explores the application of the Rectified Linear Unit (ReLU) Artificial Neural Network (ANN). {The focus is on characterizing the intricate constraints induced by FL. We opt for this neural network structure due to its versatile nature as a universal approximator and, most importantly, its reprensentability in the form of Mixed-Integer (MI) linear constraints \cite{grimstad2019relu,fischetti2018deep}.
This configuration offers the advantage in that it can seamlessly integrate with other linear constraints, including those derived from Control Lyapunov Function (CLF) or Control Barrier Function (CBF) principles \cite{li2023survey}.} Moreover, within the MPC framework, the constraints can be rigorously imposed along the entire prediction horizon without resorting to additional approximations. Notably, the MI constraints retain linearity, enabling efficient resolution through Mixed-Integer Program (MIP) solvers \cite{cplex2009v12}. To the best of the authors' knowledge, this application is novel to the literature, offering a theoretically comprehensive solution to the problem of intricate constraints associated with FL. Different from the approaches in related works \cite{hall2023differentially,greeff2021learning}, our constraint reformulation, while specifically tackling uncertainty-free problems, exhibits versatility in its applicability to multi-input systems. This adaptability is achievable through an approximation that maintains a bounded error via the ANN.

The remainder of the paper is structured as follows. Section \ref{sec:preli} presents the constrained problem and recalls the architecture of the ReLU-ANN employed. Section \ref{sec:ConstraintCharac} provides the procedure for reformulating the feedforward structure of the ReLU-ANN to MI linear constraints. {Therein, the key ingredients for the reformulation (outlined in Fig. \ref{fig:MIP_form_relu}) will be addressed and
followed by the integration in common optimization-based control settings}. Simulation studies are provided in Section \ref{sec:Sim}. Finally, Section \ref{sec:concl} concludes and addresses future directions.

\textit{Notation:} Bold lowercase letters denote vectors. For a vector $\bx\in\R^n$, $x_i$ is its $i$th component.
Comparison operators are component-wise (e.g., $\bx\leq\by$ $\Leftrightarrow$ $ x_i\leq y_i,\forall i$). $|\bx|=[|x_1|,...,|x_n|]^\top$.$\|\bx\|=\sqrt{\bx^\top\bx}$. $\boldsymbol{0}$ is the zero vector. For two integers $a<b$, $\inte{a}{b}$ denotes the set of integers inside $[a,b]$. Bold uppercase letters are matrices. $\|\bx\|_{\boldsymbol{P}}=\sqrt{\bx^\top\boldsymbol{P}\bx}$. For a matrix $\bW$, $W_{ij}$ denotes its entry on the $i$th row and $j$th column. The letters $t,t_s$ denote the time variable and the sampling time, respectively. {For a signal {$\bx(t)$},
$\bx(t_0|kt_s)$ signifies its predicted value, at the future time $t=t_0+kt_s$ upon the
information available at $t=t_0$.} For a scalar function $V(\bx)$, $\nabla V(\bx)$ is the gradient vector.

\section{Preliminaries and problem statement}
\label{sec:preli}

In this paper, we study the class of feedback linearizable systems. Formally, consider the nonlinear constrained system described by the following dynamics:
\be 
    \dot \bx = f(\bx,\bu), \text{ s.t }
    \bu \in \mathcal{U},
\label{eq:sys_0}
\ee 
where $\bx\in\R^{n_x},\bu\in\R^m$ denote the state and input vectors, respectively. $f:\R^{n_x}\times\R^m\rightarrow\R^{n_x}$ denotes the dynamics {which is assumed to be Lipschitz continuous}. For $\bu_{max}>\boldsymbol 0$, the input constraints are as follows:
\be
\mathcal{U}\triangleq
\{\bu\in\R^m:|\bu|\leq \umax\bu\}.
\label{eq:constraintU}
\ee 

\begin{figure}[bth!]
    \centering
    \includegraphics[width=0.412\textwidth]{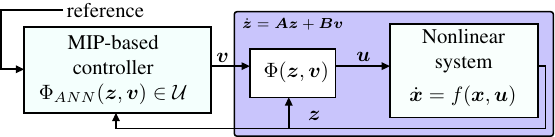}
    \caption{FL-based control with the MIP representation of the ANN.}
    \label{fig:control_scheme}
\end{figure}
Suppose that, for system \eqref{eq:sys_0}, there exist a coordinate change and an input transformation that linearize the model in closed-loop. Consequently, the system's dynamics become linear in the Brunovsky canonical form:
\be 
    \dot \bz = \boldsymbol{A}\bz + \boldsymbol{B}\bv,
    \label{eq:linearizedDyna}
\ee 
where $\boldsymbol{A},\boldsymbol{B}$ describe chains of integrators. $\boldsymbol{z}\in \R^{n_z},\bv\in\R^m$ are the new state and input variables, respectively. Note that linearization does not have to preserve the state dimension. For example, in the case of certain differentially flat systems \cite{fliess1993differentially,nicolau2017flatness}, a state prolongation is required to deduce the linearizing law (a.k.a., dynamic feedback linearization). Denote the mapping from the virtual input $\bv$ to the input $\bu$ as:
\be 
\bu = \Phi(\bz,\bv).
\label{eq:coor_Phi}
\ee 
Fig. \ref{fig:control_scheme} illustrates the role of $\Phi(\bz,\bv)$ in the purple block, creating a linear system to control in the new coordinates.
This variable change, in general, complicates the constraints \eqref{eq:constraintU} into a new set, restricting not only the new input $\bv$ but also the state $\bz$. From \eqref{eq:constraintU} and \eqref{eq:coor_Phi}, define such constraints as:
\be 
\mathcal{V}=\{(\bz,\bv):|\Phi(\bz,\bv)|\leq \umax\bu\}.
\label{eq:contraint_new_set}
\ee 

\textit{Objective:} The aim is to determine the control for the trivial system \eqref{eq:linearizedDyna} with the complicated constraint set \eqref{eq:contraint_new_set}.

\begin{rem}
\label{rem:moreConstr}
    Generally, the mapping \eqref{eq:coor_Phi} does not linearize the system in closed-loop globally. It, in fact, comes with additional operating constraints (e.g., singularity or unstable region avoidance) on the state and input, for example, $\varphi(\bz,\bv)\leq 0$. With the proposed approach, these constraints can be reformulated similarly by modifying $\mathcal{V}$ in \eqref{eq:contraint_new_set} to $\mathcal{V}'=\left\{(\bz,\bv):\Phi'(\bz,\bv)\triangleq\bbm|\Phi(\bz,\bv) |\\ \varphi(\bz,\bv) \ebm\leq \bbm \umax\bu\\0\ebm \right\}$.\QEDA
\end{rem}

{\textit{Assumption}: With such an objective, we further assume that the stabilization of \eqref{eq:linearizedDyna}
implies that of \eqref{eq:sys_0}. Dealing with unstable zero dynamics after the linearization will not be addressed in this brief paper.}

To this aim, we will show that, with 
a ReLU-ANN-based approximation
of $\Phi(\bz,\bv)$, denoted as $\Phi_{NN}(\bz,\bv)$
the constraint $(\bz,\bv)\in\mathcal{V}$ in \eqref{eq:contraint_new_set} can be inferred through MI linear constraints. This inference enables the construction of a tractable online optimization program.
To proceed, let us revisit some standard setups of the ReLU-ANN as follows.


Consider a ReLU-ANN of $K-1 $ hidden layers and, essentially, a ReLU activation function for $\by\in\R^{n_y}$:
\be 
\sigma(\by) = \max\{0,\by\} = [ \max\{0,y_1\} \,...\,\max\{0,y_{n_y}\} ]^\top.
\label{eq:max_op}
\ee 
{Denote $n_k$ number of nodes in the $k$th layer, $k\in\inte{1}{K}$ which is characterized by the weight $\bW^k\in\R^{{n_{k}\times n_{k-1}}}$ and the bias $\bb^k\in\R^{n_k}$.}
Then, the input-output relationship of the network can be calculated as follows, for  $k\in\inte{1}{K-1}$:
\be 
    \by^{k} = \sigma(\bW^k\by^{k-1}+\bb^k), 
    \by^{K}=\bW^K\by^{K-1}+\bb^K,
\label{eq:recur}
\ee 
with $\by^0$, $\by^K$ as the network's input and output, respectively.

\section{Constraint characterization with MIP 
\texorpdfstring{\\ and applications in control}{}}
\label{sec:ConstraintCharac}

In this part, we show that the constraint $(\bz,\bv)\in \mathcal{V}$ as in \eqref{eq:contraint_new_set} can be guaranteed by ensuring a set of MI linear constraints. Then, their formulation will be addressed.
The representation's application in control synthesis will be discussed subsequently.

\subsection{Constraint characterization}
\label{subsec:constr_char}
\begin{figure}[htbp!]
    \centering

\resizebox{0.475\textwidth}{!}{\input{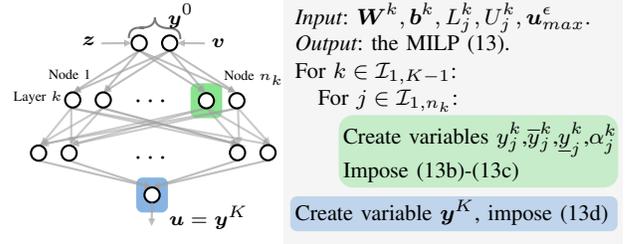}}
    \caption{The formulation of the ReLU-ANN as MILP.}
    \label{fig:MipInputCons}
\end{figure}

The constraint $(\bz,\bv)\in \mathcal{V}$ as in \eqref{eq:contraint_new_set} can be reformulated into MI linear constraints with the following proposition.
\begin{prop}
\label{prop:milc}
    Consider an approximation function of $\Phi(\bz,\bv)$ based on ReLU-ANN, called $\Phi_{NN}(\bz,\bv)$, with the network's parameters denoted as in \eqref{eq:recur}. Assume that the approximation error is bounded, i.e.,  $\exists \bep\in\R^m, \bep>\boldsymbol{0}$:
    \be 
|\Phi(\bz,\bv)-\Phi_{NN}(\bz,\bv)|\leq \bep,
\label{eq:approx_err}
    \ee 
in a compact region $(\bz,\bv)\in\mathcal{Z}\subset\R^{n_z}\times \R^{m}$. Then the constraint $(\bz,\bv)\in \mathcal{V}$ can be imposed via MI linear constraints with respect to $\bz$ and $\bv$. \QEDA
\end{prop}
\begin{proof}
The proof is twofold. First, it can be shown that the constraint \eqref{eq:contraint_new_set} can be guaranteed by imposing linear constraints on the output of the network $\Phi_{NN}(\bz,\bv)$. Namely, $|\Phi(\bz,\bv)|\leq \umax\bu$ can be implied by:
\be
|\Phi_{NN}(\bz,\bv)|\leq \umax\bu^\epsilon\triangleq\umax\bu-\bep.
\label{eq:Phi_NN_ep}
\ee 
This is because the $i$th row of $\Phi(\bz,\bv)$ is bounded as:
\begin{align}
    &|\Phi_i(\bz,\bv)|=|\Phi_i(\bz,\bv)-\Phi_{NN,i}(\bz,\bv)+\Phi_{NN,i}(\bz,\bv)|\nonumber\\
     &\leq |\Phi_i(\bz,\bv)-\Phi_{NN,i}(\bz,\bv)| + |\Phi_{NN,i}(\bz,\bv)| \\
    &\leq \epsilon_i + |\Phi_{NN,i}(\bz,\bv)|. \nonumber
\end{align}
Thus, \eqref{eq:Phi_NN_ep} implies:
$ 
|\Phi(\bz,\bv)|\leq \bep+|\Phi_{NN}(\bz,\bv)| \leq \umax\bu.
 $

Second, we show that the linear constraints \eqref{eq:Phi_NN_ep} being imposed on the ReLU-ANN can be represented via MI linear constraints.
More specifically, consider the case of a single node in the network, for some real scalar $y^{\star}$ and vector $\by^{in}$ with $\boldsymbol w,\bb$ in the appropriate dimensions:
\be 
y^{\star}=\sigma(\boldsymbol w^\top \boldsymbol y^{in}+\boldsymbol b).
\label{eq:one_perceptron}
\ee
Assume that the finite upper and lower bounds of $y^{\star}$ can be calculated, denoted respectively as $U,L$ (i.e., $L \leq y^{\star} \leq U$), then \eqref{eq:one_perceptron} can be defined implicitly as \cite{grimstad2019relu}:
\be
      \begin{cases}
   \boldsymbol w^\top \boldsymbol y^{in}+\boldsymbol b = \overline y^{\star} - \underline y^{\star} , 
    \overline y^{\star}\geq 0, \underline y^{\star} \geq 0, &\\
    \alpha \in \{0,1\},
    \overline y^{\star} \leq U \alpha, 
   \underline  y^{\star} \leq L (\alpha-1),  {y^{\star} = \overline y^{\star}},& \\
\end{cases}
\label{eq:one_perceptron_MIP}
   \ee 
where $\overline y^{\star}, \underline y^{\star}$ are real variables, and $\alpha$ is a binary variable reproducing the conditional activation (the $\max$ operation in \eqref{eq:max_op}) in each node of the network. 
Then, by applying the same reformulation for the ReLU-ANN $\Phi_{NN}(\bz,\bv)$ structured as in \eqref{eq:recur}, the constraint \eqref{eq:Phi_NN_ep} can be \textit{exactly} rewritten as:
\begin{subnumcases}
{
 \label{eq:MIP_general}
}
\by^0 = [\bz,\bv]^\top,
 k\in \inte{1}{K-1}, j\in\inte{1}{n_k},
&\label{eq:MIP_general_a}\\
    \bW^k\by^{k-1}+\bb^k=\overline \by^k-\underline \by^k,
    \overline \by^k\geq \boldsymbol 0,\underline\by^k\geq \boldsymbol 0,&\label{eq:MIP_general_b}\\
    {\alpha}_j^k \in \{0,1\},
    \overline y^k_j \leq U_j^k {\alpha}_j^k, \underline y^k_j\leq -L^k_j(1-{\alpha}_j^k),& \label{eq:MIP_general_c}\\
   \by^K=\bW^K\by^{{K}-1}+\bb^K, |\by^K| \leq \umax\bu^\epsilon \text{ as in }\eqref{eq:Phi_NN_ep},&\label{eq:MIP_general_d}
\end{subnumcases}
where, for the ReLU-ANN $\Phi_{NN}(\bz,\bv)$, {$\by^k\triangleq\overline \by^k$}, $\bW^k, \bb^k$ denote the output, weight matrix, and bias of its $k$th layer. $L^k_j,U^k_j$ are the lower and upper bound of the $j$th node in the $k$th layer. The transformation from \eqref{eq:one_perceptron} to \eqref{eq:one_perceptron_MIP} is also referred to in the literature as the ``big-M" technique.
Then, the program \eqref{eq:MIP_general} is a Mixed-Integer Linear Program (MILP) with the real variables $\by^k, \overline\by^k, \underline\by^k$ and the binary variable $\alpha_j^k$. An illustration of the formulation \eqref{eq:MIP_general} is given in Fig. \ref{fig:MipInputCons}.
\end{proof}
As noted in Fig. \ref{fig:MipInputCons}, the three key ingredients of the formulation \eqref{eq:MIP_general} include the network parameters ($\bW^k,\bb^k$), the nodes' bounds ($L_j^k,U_j^k$) and the approximation error bound ($\bep$ as in \eqref{eq:approx_err}). 
These components will be discussed subsequently.

\textit{1) The network parameters $\bW^k,\bb^k$ as in \eqref{eq:MIP_general}:}
In our setting, as the control problem is considered for a nominal model with perfect feedback (i.e., no disturbances or uncertainties), the training for $\Phi_{NN}(\bz,\bv)$ can be interpreted as a standard multivariate function fitting. For this reason, the selection of the training hyperparameters to optimize the approximation performance will not be discussed. 
In the simulation study, the ReLU-ANN will be generated by simply collecting data points from a dense grid of the space of interest and employing the toolbox from
MATLAB \cite{RegressionNeuralNetwork}.

\textit{2) The nodes' bounds $L_j^k,U_j^k$ as in \eqref{eq:MIP_general_c}:}
As well-known in the MIP community, the ``big-M" values $L_j^k$ and $U_j^k$ directly affect the solver efficiency if chosen too large, and their optimal selection, in general, still remains a topic of interest. Yet, with the focus on ReLU-ANN, several algorithms have been proposed to tighten these bounds with different levels of complexity. For a detailed survey and comparison, we send the readers to the work \cite{grimstad2019relu} and the references therein.
In the next section, the \textit{feasibility-based bound tightening} (FBBT) procedure is chosen due to its simplicity \cite{grimstad2019relu} and solver-free nature. 
With the input layer $\by^0 \triangleq [\bz,\bv]^\top\in \R^{n_z+m}$, {the calculation will be adopted from Section 4.3 in \cite{grimstad2019relu}.}





\textit{3) Error bound $\bep$ estimation in \eqref{eq:approx_err}:}
As it is well-known that ReLU-ANNs are universal function approximators \cite{yarotsky2017error}, the error bound vector $\bep$ defined in \eqref{eq:approx_err} can be optimized with a sufficiently large number of layers or nodes per layer. The sole challenge remaining is to determine and ensure this bound across a domain of interest $\mathcal{Z}\subset\R^{n_z}\times\R^m$, i.e.,
\be 
\epsilon_i=\textstyle\max_{(\bz,\bv)\in\mathcal{Z}}|\Phi_i(\bz,\bv)-\Phi_{NN,i}(\bz,\bv)|,i\in\inte{1}{m}.
\label{eq:epsilon_find}
\ee 
This approximation bound, in general, is not straightforward to obtain. 
If $\Phi(\bz,\bv)$ is continuous, the cost function in \eqref{eq:epsilon_find} is also continuous, since the ReLU-ANN $\Phi_{NN}(\bz,\bv)$ is piece-wise linear continuous. The maximizer hence exists for a {compact} domain $\mathcal{Z}$. However, since the latter is not differentiable, gradient-based techniques appear inefficient to solve this maximization problem.
In our small-scale case studies, to exploit the offline computational power, the estimation of $\bep$ will be carried out using the conventional derivative-free particle swarm algorithm.
The idea is to create numerous candidate solutions in the search space, associate them with stable virtual dynamics, and allow information exchange. In this way, the candidates can have exploratory behavior and convergence, with low computational complexity \cite{kennedy1995particle}.

{To elaborate, we will incorporate the constraint \eqref{eq:Phi_NN_ep}\footnote{{Hereinafter, when constraint \eqref{eq:Phi_NN_ep} is implemented, we mean the implementation of its MIP representation \eqref{eq:MIP_general}.}} into two effective constraint-handling frameworks: CLF-CBF and MPC.  This showcases the adaptability of the  proposed method in dealing with both stability and safety constraints, along with constraints intricately linked to model predictions.}


\subsection{Control implementation}
\label{subsec:ControlIm}
\textit{1) CLF-CBF framework:}
Given the MI linear constraint \eqref{eq:Phi_NN_ep}, one natural application is to combine it with other linear constraints, maintaining the structure of the optimization problem. Certainly, given a CLF and a CBF in the new coordinates, denoted as: $V(\bz),H(\bz):\R^{n_z}\rightarrow\R$, respectively, the online control law can be adopted as follows\cite{li2023survey}:
\begin{subequations}
	\label{eq:CLF_CBF}
	\begin{align}
&\bv^*(\bz)=\arg \textstyle\min_{\bv,\delta}J(\bv,\bz)+\delta\\
&|\Phi_{NN}(\bz,\bv)|\leq\umax\bu^\epsilon \text{ as in \eqref{eq:Phi_NN_ep}}, \label{eq:CLF_CBF_b}\\
&\nabla V(\bz)^\top(\boldsymbol{A}\bz+\boldsymbol{B}\bv)\leq -\beta V(\bz) + \delta, \ \  \delta\geq0, \label{eq:CLF_CBF_c} \\
&\nabla H(\bz)^\top(\boldsymbol{A}\bz+\boldsymbol{B}\bv)\geq -\kappa H(\bz),\label{eq:CLF_CBF_d}
	\end{align}
\end{subequations}
where $\boldsymbol{A},\boldsymbol{B}$ are from the linearized dynamics \eqref{eq:linearizedDyna}, $\kappa,\beta>0$ and $\delta\in\R$ is the relaxation variable, prioritizing the {safety certificate from \eqref{eq:CLF_CBF_d} over the tracking convergence from \eqref{eq:CLF_CBF_c}.} With regard to $J(\bv,\bz)$, by minimizing $\dot V(\bz)$, one can arrive to a mixed-integer linear program (MILP), i.e.,:
\be 
J(\bv,\bz) = \dot V(\bz) =\nabla V(\bz)^\top(\boldsymbol{A}\bz+\boldsymbol{B}\bv).
\label{eq:LP_cost}
\ee 
In another way, \eqref{eq:CLF_CBF} can be a mixed-integer quadratic program (MIQP) if $J(\bz,\bv)$ is {adapted} as:
\be 
J(\bv,\bz) = \|\bv - \boldsymbol{k}(\bz)\|^2,
\label{eq:QP_cost}
\ee 
where $\boldsymbol{k}(\bz)$ is a desired control law for $\bv$ to approach. 

\begin{rem}
    In the setting \eqref{eq:CLF_CBF}--\eqref{eq:QP_cost}, the benefit of the FL scheme is prominent, especially from the offline design view point. This is because, with the obtained integrator dynamics \eqref{eq:linearizedDyna} of which the poles have non-positive real parts, the construction of $V(\bz)$ is more apparent. The simple choice can be a quadratic CLF from the linear quadratic regulator. \QEDA
\end{rem}


{\textit{2) MPC framework:}
As the constraint \eqref{eq:Phi_NN_ep} is proposed for a general set in \eqref{eq:contraint_new_set}, its usefulness may not stand in some particular scenarios. Specifically, one may argue that
the function \eqref{eq:coor_Phi} often can be found in an affine form:
\be 
\bu = \Phi_a(\bz)\bv+\Phi_b(\bz),
\label{eq:affine_form}
\ee 
for some functions $\Phi_a(\bz),\Phi_b(\bz)$. This form renders the setup in \eqref{eq:CLF_CBF_b} unnecessary, since, given a state feedback $\bz$, such MI constraints can be replaced by a linear constraint of $\bv$:
\be 
|\Phi_a(\bz)\bv+\Phi_b(\bz)|\leq \umax\bu,
\label{eq:constr_linear}
\ee 
and \eqref{eq:CLF_CBF} become a 
QP. However, in specific applications that require model prediction, the simplicity of \eqref{eq:constr_linear} cannot be preserved, even with the advantage of the form \eqref{eq:affine_form}.
For instance, in a discrete-time MPC setting with the sample rate $t_s$, with a state feedback $\bz(t)$, the constraints \eqref{eq:constr_linear} imposed along the prediction horizon $N_p$, $k\in\inte{0}{N_p-1}$, are:
\begin{subequations}
{
\label{eq:constr_mpc_ex_full}
}
    \begin{align}
&|\Phi_a(\bz(t|kt_s))\bv(t|k t_s)+\Phi_b(\bz(t|kt_s))|\leq \umax\bu
\label{eq:constr_mpc_ex},\\
&\bz(t|(k+1)t_s) = \boldsymbol{A}_d\bz(t|kt_s) + \boldsymbol{B}_d\bv(t|kt_s),\label{eq:constr_mpc_ex2}
\end{align}
\end{subequations}
where $\boldsymbol{A}_d,\boldsymbol{B}_d$ denote the discretized model of \eqref{eq:linearizedDyna}. 
{In this formulation, the constraint \eqref{eq:constr_mpc_ex} is no longer linear with respect to the control sequence $\bv(t|kt_s)$ due to the dependence \eqref{eq:constr_mpc_ex2} involved in the multiplication $\Phi_a(\bz(t|kt_s))\bv(t|k t_s)$. Thus, even with the affine form \eqref{eq:affine_form}, the constraints \eqref{eq:constr_mpc_ex_full} are not trivial to handle \cite{kurtz1998feedback}.}
Meanwhile, with the proposed setting, we will show that the MI constraints remain linear, even when enforced over a prediction horizon as in \eqref{eq:constr_mpc_ex_full}.
}

Indeed, the constraints \eqref{eq:Phi_NN_ep} can be incorporated in an MPC, at time step $t$, as follows:
\begin{subequations}
{
\label{eq:MPC_form_miqp}
}
    \begin{align}
        &\bv^*={\arg}\textstyle\min_{\bv(t|kt_s)}\sum_{k=0}^{N_p-1}\ell(\bz(t|kt_s),\bv(t|kt_s))\label{eq:MPC_form_miqp_a}\\
       & |\Phi_{NN}(\bz(t|kt_s),\bv(t|kt_s))|\leq\umax\bu^\epsilon ,\label{eq:MPC_form_miqp_b}\\
       &\bz(t|(k+1)t_s)=\boldsymbol A_d \bz(t|kt_s)+\boldsymbol B_d\bv(t|kt_s),\label{eq:MPC_form_miqp_c}\\
       &\bz(t|kt_s)\in\mathcal{X}_z,k\in\inte{0}{N_p-1},\label{eq:MPC_form_miqp_d}
    \end{align}
\end{subequations}
where $\boldsymbol{A}_d,\boldsymbol{B}_d$ denote the discretized model of \eqref{eq:linearizedDyna} with sampling time $t_s$, $\mathcal{X}_z $ is the polytopic state constraint, and $\ell(\bz,\bv)$ is a chosen cost function.
While the constraints \eqref{eq:MPC_form_miqp_c}-\eqref{eq:MPC_form_miqp_d} are linear, from Proposition \ref{prop:milc}, the MI constraint \eqref{eq:MPC_form_miqp_b} is also linear for the variables $\bv(t|kt_s),\bz(t|kt_s)$.
Thus, with a standard quadratic cost $\ell(\bz,\bv)$, one can formulate a MIQP-based MPC with \eqref{eq:MPC_form_miqp}. 
To show the efficacy of the methods, we will examine numerical examples in the next section.

\section{Simulation study}
\label{sec:Sim}
For simplicity, hereinafter, for all ReLU-ANN approximations with $K-1$ hidden layers, each layer will have equally $n_k={\bar{n}}$ nodes. The training data will be sampled from a hyperbox noted as $\mathcal{T}$ with $n_s$ samples in each axis. The training is conducted with the method \texttt{fitrnet} from \cite{RegressionNeuralNetwork} with default hyperparameters. Numerical specifications for the characterization are given in Table \ref{tab:specs}. {The training time (TT) is also reported therein.} 
All the MIPs were solved with the Yalmip toolbox \cite{Lofberg2004} and CPLEX solver \cite{cplex2009v12}.
Next, the simulation studies will be presented, followed by the discussion.

\subsection{Nonlinear mass-spring damper system (MSD)}
\label{subsec:MSD_sim}

\begin{figure}[htb]
    \centering
    \includegraphics[width=0.465\textwidth]{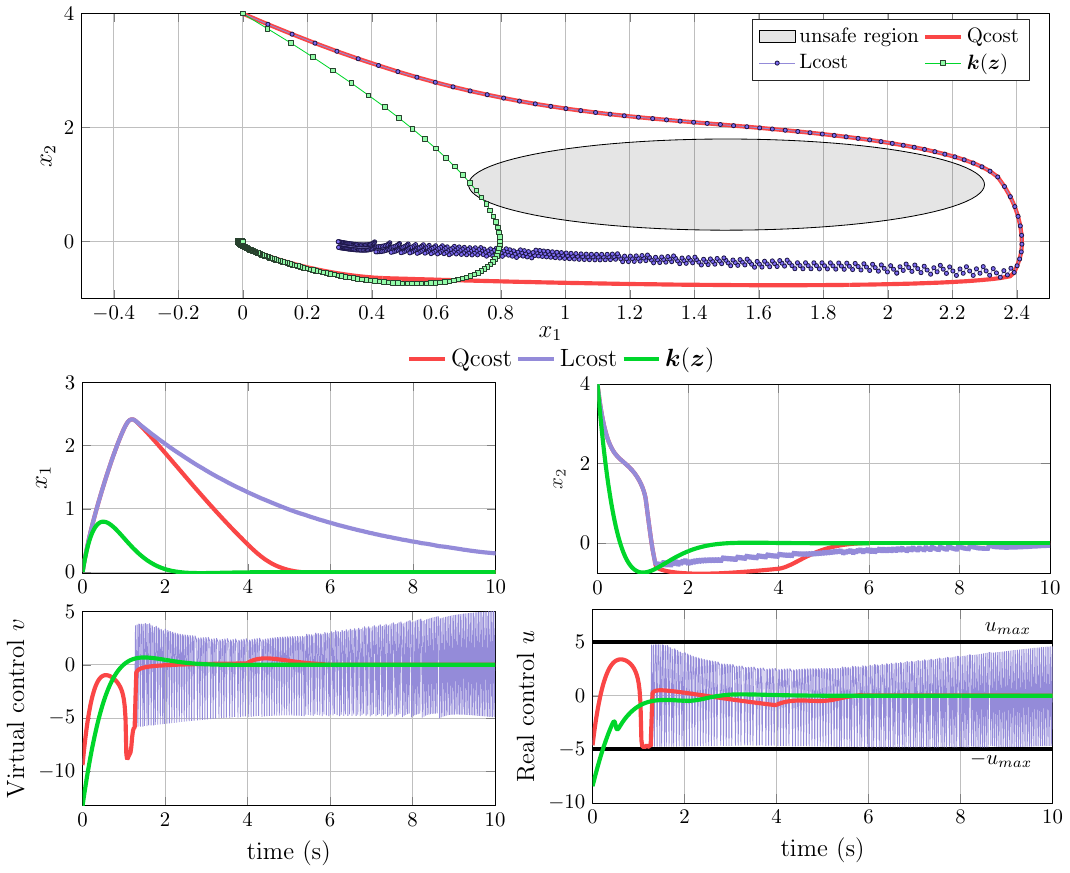}
    \caption{{Stabilization with ANN-based constraint characterization.}}
    \label{fig:simMSD}
\end{figure}
Consider the dynamics \cite{romdlony2016stabilization}:
\be 
\dot x_1=x_2,\dot x_2=-s(x_2)-x_1+u,
\label{eq:MSD_model}
\ee 
where $x_1,x_2$ represent the displacement and the velocity of the system, respectively, and $u$ is the constrained input signal with $|u|\leq\umax u=5$. The damping is described by the Stribeck friction model $s(x_2)=(0.8+0.2e^{-100|x_2|})\tanh(10x_2)+x_2$.
The model can be written in the form of \eqref{eq:linearizedDyna} with $\bz=[z_1,z_2]^\top=[x_1,x_2]^\top$, $\boldsymbol{A}=\begin{bsmallmatrix}
    0&1 \\ 0&0
\end{bsmallmatrix}$, $\boldsymbol{B}=[0,1]^\top$, $\Phi(\bz,v)=v+s(z_2)+z_1$ and the input constraint now becomes $
|\Phi(\bz,v)|\leq \umax u
$ which will be approximated and converted to MI constraints as in Section \ref{subsec:constr_char}, with training points sampled from $ \mathcal T=\{|\bz|\leq 5, |v|\leq 10\}$.
\begin{table}[bt!]
    \centering
    \caption{Numerical specifications}
    \begin{tabular}{|c|c|c|c|c|c|c|c|c|}
    \hline
        System & $K$ &${\bar{n}}$   & $\bep$ in \eqref{eq:approx_err} &$\kappa $&  $\beta$&$n_s$ & {TT(s)}\\ \hline
       MSD \eqref{eq:MSD_model} & 4 & 20  &0.2740 & 4.0& 0.001&30      &{9.43} \\  \hline
       Quad-1D \eqref{eq:quad1D_nonlinear}  & 4 & 10  & 0.0322& $\times$ &$\times$ &35  & {2.19}\\ \hline
    \end{tabular}
    \label{tab:specs}
\end{table}

To examine the setup in \eqref{eq:CLF_CBF}, consider an unsafe region 
$\mathcal O =\{\bz:H(\bz)\leq 0\}$ with $H(\bz)\triangleq\|\bz-\bz_o\|- r_o$, $\bz_o=[1.5,1]^\top,r_o=0.8$.
For control, the program \eqref{eq:CLF_CBF} will be employed with both the linear cost (Lcost) as in \eqref{eq:LP_cost} and quadratic cost (Qcost) as in \eqref{eq:QP_cost}. The CLF can be chosen as a standard quadratic function $V(\bz)=\|\bz\|_{\boldsymbol{P}}^2 $, $\boldsymbol{P}=\begin{bsmallmatrix}
	 4.58 & 10 \\ 10 &45.83
\end{bsmallmatrix}$ and $\boldsymbol k(\bz)=-[4.47,\,3.37]\bz$.
With $\bz(0)=[0,4]^\top$, the simulation results were given in Fig. \ref{fig:simMSD} with the three controllers: Lcost, Qcost for \eqref{eq:CLF_CBF} and $\boldsymbol{k}(\bz)$ given previously.

\begin{figure}[hbt]
    \centering    \includegraphics[width=0.46\textwidth]{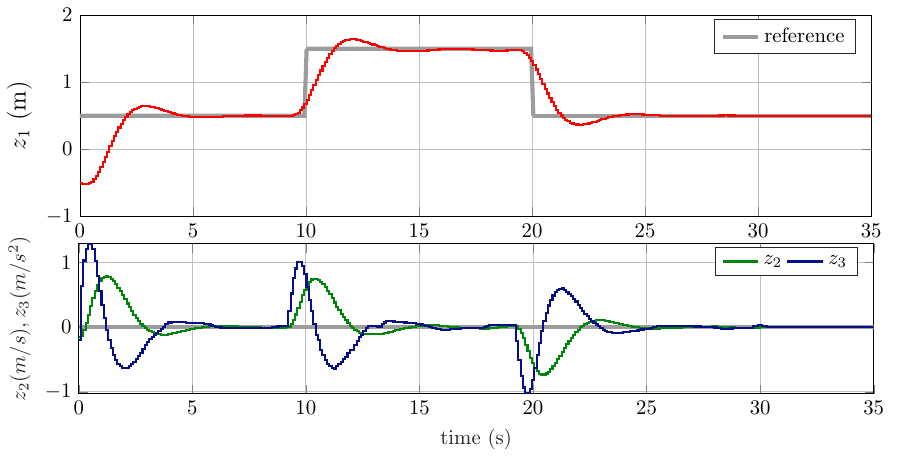}
    \caption{Trajectory tracking with the MPC \eqref{eq:MPC_form_miqp} for \eqref{eq:quad1D_nonlinear}.}
    \label{fig:Quad1D_state}
\end{figure}

\subsection{Horizontal 1D quadrotor (Quad-1D)}
\label{subsec:quad1D}
To validate the applicability of the MPC setup in \eqref{eq:MPC_form_miqp}, we examine the horizontal 1-D quadcopter model \cite{greeff2020exploiting}:
\be \ddot x=\Gamma \sin\theta - \gamma \dot x,\, \dot \theta=\tau^{-1}(u-\theta),
\label{eq:quad1D_nonlinear}
\ee 
where $x$ is the horizontal displacement, $\theta$ is the pitch angle. $u$ is the commanded pitch angle and constrained as $|u|\leq \umax u= 0.1745$ (rad). The model parameters are $\Gamma=10,\gamma=0.3,\tau=0.2.$
With $z_1=x$, the nonlinear model \eqref{eq:quad1D_nonlinear} can be linearized to:
$
\dddot z_1 = v, 
$
with the mapping:
\be 
\Phi(\bz,v)=
\tau(v+\gamma z_3)/(\Gamma\sqrt{1-\zeta^2(\bz)} 
)+\sin^{-1}\zeta(\bz),
\ee 
where $\bz=[z_1,z_2,z_3] = [x,\dot x, \ddot x]$,$\zeta(\bz)=\Gamma^{-1}(z_3+\gamma z_2)$. The new constraint hence yields $|\Phi(\bz,v)|\leq \umax u$ which is then approximated with samples from $ \mathcal T=\{|\bz|\leq 5, |v|\leq 15\}$.
\begin{figure}[t]
    \centering
    \includegraphics[width=0.47\textwidth]{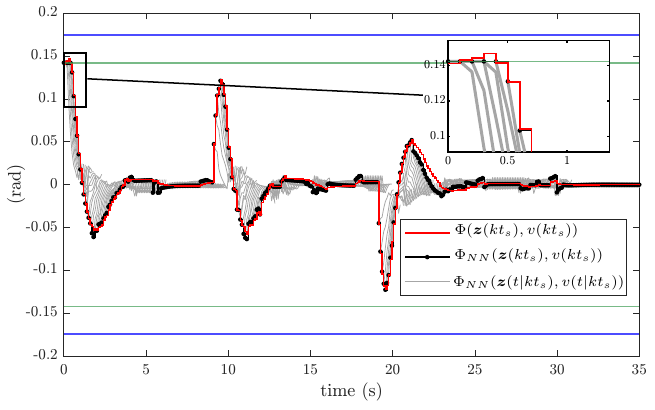}
    \caption{{The input $u=\Phi(\cdot)$, the approximation $\Phi_{NN}(\cdot)$ and its predicted trajectories}.$|u|=\pm\umax u$ (blue) and $u=\pm\umax u^\epsilon$ (green).}
    \label{fig:Quad1D_input}
\end{figure}

To construct the MPC \eqref{eq:MPC_form_miqp}, the linearized system
is rewritten in the form of
\eqref{eq:linearizedDyna} and discretized with the forth-order Runge-Kutta method, sampled at $t_s=100$ms. The prediction horizon is $N_p=10$ steps, the time-varying reference $\bz_{ref}$ is depicted in gray on Fig. \ref{fig:Quad1D_state} and the cost function $\ell(\cdot)$ as in \eqref{eq:MPC_form_miqp_a} is chosen as: $
    \ell(\bz,v) = \|\bz-\bz_{ref}\|_{\boldsymbol{Q}}^2+\|v\|_{\boldsymbol{R}}^2,
$ with $\boldsymbol{Q}=\begin{bsmallmatrix}
    10 & 0 & 0\\ 0 &1& 0 \\ 0&0&1
\end{bsmallmatrix}$, $\boldsymbol{R}=0.0175$, $\bz(0)=\begin{bsmallmatrix}-0.5&-0.15&-0.2\end{bsmallmatrix}^\top$. The system's trajectory is given in {Fig. \ref{fig:Quad1D_input} and \ref{fig:Quad1D_state}}, {while the online computation time in both examples is reported in Table \ref{tab:computationtime} with its min, max and mean value.}

\begin{table}[bh]
  \centering
  \caption{Computation time of the simulations (ms).}
    \begin{tabular}{|c|c|c|c|c|}
    \hline
     System    & Controller & min  & max  & mean \\
    \hline
    \multirow{3}[0]{*}{MSD \eqref{eq:MSD_model}} & Lcost   &   $20.96$  &    $34.52 $ &   $25.15$ \\
\cline{2-5}         & Qcost   &   $20.62$    &  $30.81$    & $23.83$ \\
\cline{2-5}         & $\boldsymbol k(\bz) $&    $8\mathrm{e}{-4}$  &   $0.29$   & $2.21\mathrm{e}{-3}$ \\
    \hline
    Quad-1D \eqref{eq:quad1D_nonlinear}& MPC $(N_p=10)$&$  42.76 $   &   $5.96\mathrm{e}{+4}$   &  $ 1.69\mathrm{e}{+3}$\\
    \hline
    \end{tabular}%
  \label{tab:computationtime}%
\end{table}%
\vspace{-0.5cm}

\subsection{Discussion}
\label{subsec:discuss}

In general, in both investigated systems, the proposed setting is compatible with standard optimization-based control framework, achieving stability, safety (avoidance of the unsafe region), and most importantly, constraint satisfactions, even for forecasted trajectories. 

\begin{figure}[hpt]
    \centering
    \includegraphics[width=0.475\textwidth]{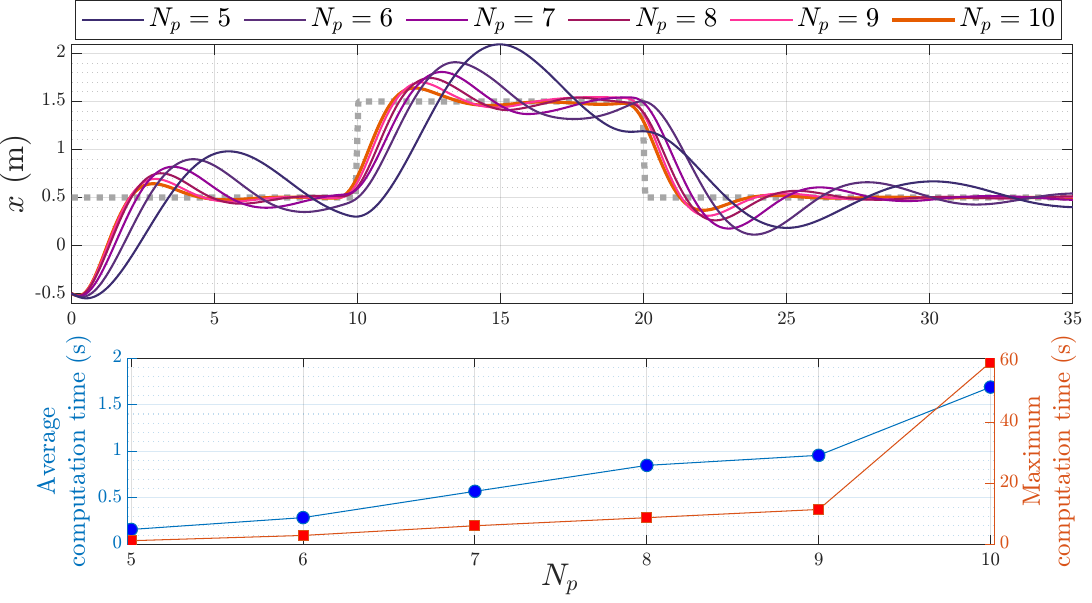}
    \caption{The displacement $x(t)$ and the computational cost w.r.t $N_p$.}
    \label{fig:Quad1D_state_Comptime}
\end{figure}

For system \eqref{eq:MSD_model}, from Table \ref{tab:computationtime}, it is noteworthy that both Lcost and Qcost controllers can be implemented with relatively low computation time (from 20-40ms). The only shortcoming is the discontinuity of the Lcost controller, which is caused by the non-smooth characteristics of the constraints \eqref{eq:CLF_CBF_b}--\eqref{eq:CLF_CBF_d} in combination with the linear cost. Yet, the problem is not present for the Qcost controller, and interestingly, the computation cost is even lower as compared to that of Lcost. 
Stable tracking can also be found in the example of the 1D quadrotor \eqref{eq:quad1D_nonlinear}, where a time-varying trajectory can be tracked with the advantage of anticipation from MPC (see Fig. \ref{fig:Quad1D_state}). Given the error bound as in \eqref{eq:approx_err}, the input constraint is guaranteed by imposing $|\Phi_{NN}(\cdot)|\leq \umax u^\epsilon$.
It is crucial to emphasize that in our approach, neither the predicted constraint nor the model underwent approximation through state feedback, as observed in \cite{kurtz1998feedback,nevistic1994feasible}.
Hence, the constraints were {rigorously} enforced over the prediction horizon for the nominal dynamics.
This advancement, from a theoretical standpoint, paves the way for constrained FL-based control design with rigorously guaranteed stability. One potential resolution involves adopting the standard MPC axioms \cite{mayne2000constrained}, incorporating a stabilizing selection of parameters $N_p,\mathcal{X}_z$ and $\ell(\cdot)$ in \eqref{eq:MPC_form_miqp}.
This reiterates the necessity of delving deeper into understanding the proposed MI linear constraint \eqref{eq:Phi_NN_ep} on a case-by-case basis.

Finally, although satisfactory results and general
development were found, the real-time implementation with MPC is evidently still hindered by the
solving of the MIQP \eqref{eq:MPC_form_miqp} (see Fig. \ref{fig:Quad1D_state_Comptime}), especially when the number of binary variables grows with the number of neurons and the prediction horizon\footnote{{
Each time \eqref{eq:MPC_form_miqp_b} is imposed, $\textstyle\sum_{k=1}^{K-1}n_k$ binary variables are used. Hence, over $N_p$ steps, the number of binary variables is $N_p\textstyle\sum_{k=1}^{K-1}n_k$.
}} $\left(N_p\textstyle\sum_{k=1}^{K-1}n_k\right)$.
{This implementational shortcoming becomes even more evident when multi-input systems are considered, since the linearizing mapping will certainly require a larger network.}
For the method to be real-time capable, future work concerns the simplification of the constraints and an optimal choice of the network parameters.

\section{Conclusion and outlook}
\label{sec:concl}
This work addressed the feedback linearization control problem with the convoluted constraints caused by the linearizing coordinate change. 
The constraint characterization includes approximating the mapping by a rectified linear unit artificial neural network and reformulating it into the equivalent mixed-integer linear constraints. The applicability and validity of the proposed method were shown via simulation tests with the control Lyapunov function-control barrier function, and model predictive control strategies.
{Future theoretical directions
focus on the investigation of the new constraints and their adaptation within the standard linear control design, showcasing the advantage with comparative study.} {Meanwhile, practical extensions will target optimization in the design parameters and the method's applications to multi-input systems}.



\end{document}